\documentclass[prc,aps,eqsecnum,epsf,amssymb]{revtex4}
\usepackage{graphicx}

\newcommand{\bmq}{{\mbox{\boldmath $q$}}}

\begin{document}
\preprint {WIS/23/05-Sept-DPP}
\date{\today}
\title{On the relation between nuclear and nucleon Structure Functions 
and their moments}
\author{A.S. Rinat and M.F. Taragin}
\address{Weizmann Institute of Science, Department of Particle Physics,
Rehovot 76100, Israel}

\begin{abstract}

Calculations of nuclear Structure Functions (SF) $F_k^A(x,Q^2)$ routinely exploit 
a generalized convolution, involving the SF for nucleons $F_k^N$ and the linking SF 
$f^{PN,A}$ of a fictitious nucleus, composed of point-particles, with the 
latter usually expressed in terms of hadronic degrees of freedom. For finite 
$Q^2$ the approach seemed to be lacking a solid justification and the same 
is the case for recently proposed, effective nuclear parton distribution functions 
(pdf), which exactly reproduce the above-mentioned hadronically computed $F_k^A$. 
Many years ago Jaffe and West proved the above convolution in the Plane Wave 
Impulse Approximation (PWIA) for the nuclear components in the convolution. In the 
present note we extend the above proof to include classes of nuclear Final State 
Interactions (FSI). One and the same function appears to relate parton distribution 
functions (pdf) in nuclei and nucleons, and SF for nuclear targets and for nucleons. 
That relation is the previously conjectured one,with an entirely 
different interpretation of $f^{PN,A}$. We conclude with an extensive 
analysis of moments of nuclear SF based on the generalized convolution. 
Characteristics of those moments are shown to be quite similar to the same 
for a nucleon. We conclude that the above evidences asymptotic freedom of 
a nucleon in a medium and not of a composite nucleus. 

\end{abstract}

\maketitle

\section{Introduction.}

This note concerns two related topics. The first is a generalized convolution, 
involving Structure Functions (SF) $F_k^A$ and $F_k^N$, which compose 
cross sections for inclusive scattering of unpolarized leptons from composite 
targets $A$ and for a nucleon. The second one deals with implications of the 
above for moments of $F_k^A$.

Standard approaches employing hadronic degrees of freedom have used
generalized convolutions of the form 
\begin{mathletters}
\label{a1}
\begin{eqnarray}
 F^A&=&f^A*F^N
\label{a1a}\\  
 F^A_k(x,Q^2)&=&\sum_{a}\int_x^A\frac {dz}{z^{2-k}} f^{a,A}(z,Q^2)     
     F_k^a \bigg (\frac {x}{z},Q^2\bigg )       
\label{a1b}\\                                         
         &\approx&\int_x^A\frac {dz}{z^{2-k}} f^{PN,A}(z,Q^2)
     F_k^{\langle N \rangle}\bigg (\frac {x}{z},Q^2\bigg )
\label{a1c}\\ 
F_k^{\langle N \rangle}=\frac {ZF_k^p +NF_k^n}{A}
&=& \frac{1}{2}\bigg [1-\frac {\delta N}{A}\bigg ]F_k^p+
\frac{1}{2}\bigg [1+\frac {\delta N}{A}\bigg ]F_k^n
\label{a1d}   
\end{eqnarray}
\end{mathletters}
The involved SF depend on the squared 4-momentum transfer 
$q^2=-Q^2=-(|\bmq|^2-\nu^2)$ and on the Bjorken variable $x$ in terms of 
the nucleon mass $M$ with support $0 \le x=Q^2/2M\nu \le M_A/M \approx A$. 

Eq. (\ref{a1b}) decomposes $F_k^A$ into contributions from various constituents 
$'a'$, such as nucleons, virtual bosons, etc. For the kinematic region of our main 
interest, $x\gtrsim 0.2$, it suffices to retain only nucleons, or more precisely, 
the averaged nucleon with  SF $F_k^{\langle N \rangle}$, Eq. (\ref{a1d}), obtained 
by weighting $F_k^{p,n}$ with $Z,N$: $\delta N/A$ is the relative neutron 
excess.  

Within the framework of hadron dynamics, the convolution (\ref{a1c}) can be
proven in the PWIA \cite{pace}. In that approximation the linking function $f$ 
in Eq. (\ref{a1c}), which in general is the SF of a fictitious nucleus composed 
of point-nucleons, is approximated by $f^{PN,A} \to f^{PWIA}$, with the latter
related to the spectral function of the knocked-out nucleon in the target 
\cite{x1}. For finite $Q^2$, Eq. (\ref{a1c}) stood as a conjecture. 

The same is the case for an alternative, non-perturbative Gersch-Rodriguez-
Smith (GRS) approach \cite{gr1}, which has originally been formulated for 
a non-relativistic system of point-particles \cite{grs}. It has subsequently 
been extended to systems of composite constituents, such as quantum gases 
and liquids H$_2$, D$_2$, He etc. Since the energy scales for electronic, 
rotation-vibration modes, etc. differ appreciably, the Born-Oppenheimer 
approximation applies. As a consequence the SF (or $'$linear response$'$) 
of the composite system, is accurately given as a repeated regular convolution 
(\ref{a1a}), involving SF of the translation of the centers of mass of inert 
molecules and of internal modes of each molecule \cite{yk} 
\begin{eqnarray}
F^{qu\, gas}(|\bmq|,\nu)=\int d\nu_1 F^{trsl}(|\bmq|,\nu-\nu_1)
&&\int d\nu_2 F^{rot}(|\bmq|,\nu_1-\nu_2) 
\nonumber\\
&&*\int d\nu_3 F^{vibr}(|\bmq|,\nu_2-\nu_3)*....
\label{a2}
\end{eqnarray}
The next step in the development
has been a covariant generalization of the above GRS theory, first for the 
SF of a system of point-particles, i.e. for $f$ in (\ref{a1c}) \cite{gr2}. 
For increasing $Q^2$, internal degrees of freedom need  ultimately to be 
included through $F_k^{\langle N \rangle}$, as described by the generalized 
convolution (\ref{a1c}). 

It stands to reason, that in general the convolution Eq. (\ref{a1c}) 
for a composite nucleus rests on different energy scales for the 
participating modes. In fact, Eq. (\ref{a1c}) was proven for a model 
with quarks clustered in nucleons, where the energy scale for internal 
excitations is much in excess of the same for $NN$ forces \cite{gr1}. For 
higher, but not asymptotic $Q^2$, it seemed difficult to derive a 
covariant version and Eq, (\ref{a1c}) has been considered a conjecture. 

Calculations were  based on data for $F_2^p$ and on some adopted 
$F_k^n$ \cite{rt2}, such that a calculation of $F_k^A$ amounts to
the same of $f^{GRS}$. The latter can be evaluated using purely hadronic 
notions, such as single-nucleon spectral functions, nuclear density 
matrices of various orders, $NN$ forward scattering amplitudes, etc. 
Support for the validity of Eq. (\ref{a1c}) came mainly from the 
satisfactory description of a large body of inclusive scattering cross 
sections data for $Q^2\gtrsim Q_0^2 \approx 2.5\,$GeV$^2$ \cite{rtval,rt,vkr}.

Also for later reference, we mention that Eq. (\ref{a1c}) has its 
deficiencies. For example, $F_k^{\langle N \rangle}$ is taken 
to be the SF of a free averaged nucleon, which generally is off its
mass-shell. In addition, Eq. (\ref{a1c}) lacks explicit spin, iso-spin 
structure and in particular $f$ is usually computed from spin, iso-spin
averaged input.

Next we recall an alternative representation of nuclear SF, which uses 
nuclear parton distribution functions $q_i^A(x,Q^2)$ for finite $Q^2$, 
which has to be computed from their nucleonic analogs  $q_i(x,Q^2)$. 
Those nuclear pdf are effective ones: we do not aim for an underlying 
theory, and in particular not for accounting of $Q^2$ 
dependence, compatible with evolution from a scale $Q_0^2$. The only 
requirement is the exact reproduction of $F_2^A(x,Q^2)$, as 
computed in the hadronic representation (\ref{a1c}). 

The above requirement is nowhere sufficient to determine those pdf, and the
apparent freedom is exploited by making two deliberate choices \cite{rt3}. 
An inessential one assumes $F_2^A$ to be the same combination of 
nuclear parton dfs, as $F_2^{\langle N \rangle}$ is of nucleon ones, 
thus (for clarity we drop the $x,Q^2$ dependence in arguments)
\begin{mathletters}                                    
\label{a3}                                             
\begin{eqnarray}                                       
F_2^{\langle N \rangle}={\sum_i}a_ixq_i&=&
\frac {5x}{18}\bigg [u_v+d_v+2{\bar u}+2{\bar d}+
\frac {4}{5}s-\frac{3\delta N}{5A}
(u_v-d_v+2{\bar u}-2{\bar d})\bigg ],
\label{a3a}\\
F_2^A\equiv{\sum_i}a_ixq_i^A&=&
\frac{5x}{18}\bigg [u_v^A+d_v^A+2{\bar u}^A+2{\bar d}^A+
\frac{4}{5}s^A-\frac{3\delta N}{5A}(u_v^A-d_v^A+2{\bar u}^A-2{\bar d}^A)\bigg ] 
\label{a3b}                                            
\end{eqnarray}                                         
\end{mathletters}  
Next we chose to relate nuclear pdf of given species $i$ to its analog 
for the averaged nucleon $\langle N \rangle$, in precisely the same way as 
the hadronic representation (\ref{a1c}) links nuclear and nucleon SF, thus
\begin{mathletters}
\label{a4}
\begin{eqnarray}
q_{i/A}(x,Q^2)&=&\sum_{a} \int_0^A dz f_{a/A}(z,Q^2) q_{i/a} 
\bigg (\frac {x}{z},Q^2\bigg )
\label{a4a}\\
     &\approx &\int_0^A dz f_{N/A}(x,Q^2)q_{i/N}\bigg (\frac {x}{z},Q^2\bigg ),
\label{a4b}
\end{eqnarray}
\end{mathletters}
Eq. (\ref{a4b}) does not mix flavors and uses a single linking function 
$f_{N/A}=f^{PN,A}$, independent of the species, whether valence, sea quarks 
or gluons. By construction the computed nuclear SF $F_2^A$ ({\ref{a3b}) 
in the parton representation ({\ref{a4b}) are identical to their hadronic 
analog ({\ref{a1c}), provided the same input is used. 
In practice the input $F_2^n$ for the two differs (see Ref. \cite{rt3} 
for a discussion). Between parenthesis we add, that being the same SF as 
in Eq. (\ref{a1c}), $f$ carries along the above-mentioned deficiencies.

For both the hadron and pdf representations of $F_2^A$, there seemed to be 
missing a proof of Eqs. (\ref{a1c}), as well an estimate of the lower limit 
$Q_0^2$, beyond which Eq. (\ref{a3b}) is approximately valid. However, 
we recently stumbled upon 20 year old papers by Jaffe and West, which 
contain the basics of the desired proof \cite{jaffe,west}. Judging from
the lack of citations, even cognoscenti apparently overlooked or forgot 
the above papers, possibly because those were published in the proceedings 
of a summer school and of an AIP meeting. In the above publications the 
generalized convolution is derived, using a parton model as well as 
pQCD, both in the special case of the PWIA. In the following we generalize 
their results to include nuclear FSI.

Since the article of Jaffe is fairly self-contained, it will suffice to only 
cite some essentials, and in particular the central relation between forward 
$\gamma$-target scattering amplitudes and pdf. We then show that, although 
the inclusion of $general$ FSI usually spoils their accommodation in a 
convolution for $F_k^A$ \cite{jaffe}, this is not the case for some nuclear 
FSI not involving partons. The above holds for instance for the Distorted 
Wave Impulse Approximation (DWIA) in the form given in Ref. \cite{rj}. The 
same is the case for the GRS version for finite, relatively large $Q^2$ 
and those a $f^{Pn,A}$ are  precisely the ones contained in Eqs. (\ref{a4b}) 
and (\ref{a1c}), the above indeed completes the proof for, what previously 
was called a conjecture.

We conclude this note with the analysis of data on moments of high-$Q^2$ 
nuclear SF and  present pQCD results for the above as done in the past for
a proton.

\section{Derivation of nuclear parton distribution functions.}

We start with a proton and consider the forward scattering amplitude 
(fsa) $a(\gamma p$) as a two-step process, where the proton emits a
quark, which in turn absorbs the virtual photon (Fig. 1). That 
amplitude can be evaluated, given an expression for the current in terms of 
parton fields. For instance in a model with free parton fields, the 
result is \cite{jaffe}
\begin{mathletters}
\label{a5}
\begin{eqnarray}
F_2^p(x,Q^2)&=&x\sum_i e_i^2q_i(x,Q^2)
\label{a5a}\\
q_i(x,Q^2)  &=& \int \frac{d^4k}{(2\pi)^4}
\delta \bigg (x-\frac{k.q}{p.q}  \bigg )\chi_i(k,p)
\label{a5b}\\
q_i^{scal}(x)={\lim_{Q^2\to\infty}} q_i(x,Q^2)&=& \int\frac {d^4k}
{(2\pi)^4}\delta \bigg (x-\frac{k^+}{p^{+}}\bigg )\chi_i(k,p),
\label{a5c}
\end{eqnarray}
\end{mathletters}
with the sum in Eq. (\ref{a5a}) over quarks with charge $e_i$. 
$\chi_i=\chi_{i/p}$ in Eq. ({\ref{a5c}) is the fsa $a(q_ip)$ in Fig. 1 . 
Above one neglects spin and color: their inclusion is straight-forwarded, 
and is immaterial for the reasoning. To lowest order, i.e. in the 
PWIA, Eq. (\ref{a5b}) is proportional to, what in nuclear physics parlance 
is called, the spectral function of a parton $i$ in the $p$. The 
$\delta$-function in the integrand of Eq. (\ref{a5b}) selects the momentum 
fraction $x$ of the quark in the proton as determined by the 4-momenta 
$k,p,q$ of the quark, proton and virtual photon and the integrand in 
Eq. (\ref{a5b} holds
for finite $Q^2$. Eq. (\ref{a5c}) is the Bjorken limit of the
above, in which case the argument of the $\delta$ function can be expressed 
in terms of the dominant light-cone components $k^+,p^+$: the resulting 
$q_i, F_2^p$ are pdf and SF in the scaling limit and depend only on $x$.

Of an entirely different nature is the $Q^2$-dependence generated by $'$FSI$'$ 
beyond the PWIA, coming from quarks which emit gluons, from gluon pair-production, 
triple gluon coupling, etc. Those add ln($Q^2)$ and $[1/Q^2]^n$ corrections to 
the above scaling limits for pdf and SF. For the present purpose it is 
irrelevant whether those ultimately derive from the Operator Product Expansion 
(OPE), or are calculated  in pQCD by evolution. 

Much of the above for a $p$ target, a neutron, or  averaged $N$, holds 
also for a general target $A$: One can copy Eqs. (\ref{a5a}), (\ref{a5b}) and 
(\ref{a5c}) replacing $p(N)$ by a composite target. However, it is awkward 
to deal with the spectral function of a parton in a nucleus, as is the fsa 
$\chi_{i/A}$ in the PWIA. 

A more natural way is the evaluation of that amplitude upon insertion 
of an intermediate set of states for free nucleons and a fully 
interacting daughter nucleus. The product of the fsa $a(\gamma q_i)$ 
and $a(q_i N)$ is subsequently integrated over the intermediate momentum 
in order to form $a(\gamma N)\propto\, F^N$. The result, illustrated in 
Fig. 2, amounts to the following relation between the three involved fsa
\begin{eqnarray}
\chi_{i/A}(k,P)=\sum_a\int \frac {d^4p}
{(2\pi)^4}\chi_{i/a}(k,p)\chi_{a/A}(p,P),
\label{a6}
\end{eqnarray}
where the two sub-amplitudes for $\gamma q$ and $N$-Sp are in the PWIA. The
fsa  $a(N$-Sp) in the PWIA is now related to the familiar spectral function
of a nucleon in the target. 

As in Eq. (\ref{a5b}) for a $p$, one now projects out of each fsa the 
appropriate pdf, Eq. (\ref{a6}) is converted to  
\begin{mathletters}
\begin{eqnarray}
\label{a7}
q_{i/A}(x)&=&\sum_a \int_x^A dz \int {dp_0}^2 f_{a/A}(z;p_0^2)
q_{i/a}\bigg (\frac {x}{z};p_0^2 \bigg )
\label{a7a}\\
&\approx &\int_x^A dz f_{N/A}(z)
q_{i/N} \bigg ( \frac {x}{z} \bigg )
\label{a7b}
\end{eqnarray}
\end{mathletters}
Again Eq. (\ref{a7a}) relates to several constituents/clusters , all of 
which may be off their mass-shell ($p_0^2\ne M_a^2$), while in Eq. 
(\ref{a7b}) one only retains nucleons, and in addition disregards 
those off-shell 
effects. Eq. (\ref{a7b}) is clearly Eq. (\ref{a4b}) in the Bjorken limit.

Next, upon inclusion of gluon emissions from quarks, nuclear pdf acquire 
$Q^2$-dependence, changing Eqs. (\ref{a7a}) and (\ref{a7b}) into
\begin{mathletters}
\begin{eqnarray}
\label{a8}
q_{i/A}(x,Q^2)&=&\sum_a \int_x^A dz \int {dp_0}^2 f_{a/A}(z,Q^2;p_0^2)
q_{i/a}\bigg (\frac {x}{z};Q^2;p_0^2 \bigg )
\label{a8a}\\
&\approx &\int_x^A dz f_{N/A}(z,Q^2)
q_{i/N} \bigg ( \frac {x}{z},Q^2 \bigg )
\label{a8b}
\end{eqnarray}
\end{mathletters}
Above $f_{N/A}$ is the df of nucleons in the nucleus in the PWIA, while 
$q_{i/N}$ are pdf beyond their scaling limit. Now just as gluon effects  
may be viewed as FSI on the fsa $a(\gamma q_i)$ in the in the scaling 
limit, one should consider FSI pertinent to the nuclear part.

As emphasized by Jaffe, most classes of those FSI cannot be accommodated in a 
generalized convolution. However, the above does not hold for selected, 
nuclear FSI, generated by the interaction between the above-assumed 
free $N$ and the spectator nucleus. An illustrative example is a ladder of 
$N$-spectator collisions, which turn the PWIA into the DWIA (Fig. 3). 
The same holds for a description in the alternative, non-perturbative GRS 
theory for FSI: $f_{N/A}\to f^{PN,A}\to f^{GRS}$, which leads to the GRS 
version of Eq. (\ref{a1c}) \cite{gr1,gr2}.  
 
In a last step one takes the proper combinations (\ref{a3a}), (\ref{a3b}) of 
nucleon, respectively nuclear pdf, and obtains 
\begin{eqnarray}
F_2^A(x,Q^2)=\int_x^A dz f^{PN,A}(z,Q^2)
          F_2^{\langle N \rangle} \bigg (\frac {x}{z},Q^2 \bigg )
\label{a9}
\end{eqnarray}
Eqs. (\ref{a8b}), (\ref{a9}) are manifestly the same as Eqs. (\ref{a4b}),
(\ref{a1c}), but Eq. (\ref{a5c}) is a $choice$, whereas 
Eq. (\ref{a8b}) is the result of a $derivation$. Just as for the
descriptions outlined in Sections I, II, one deals with one, 
species-independent $f_{N/A}$, which relates df of partons in nuclei and 
nucleons without flavor mixing. We recall, that the above 
correspondence holds for the two discussed approaches, in which quite 
similar approximations have been applied, e.g. the use of averaged 
spin-isospin observables and the neglect of off-shell effects. 
Finally, not all even $'$purely$'$ nuclear FSI components can be
accommodated in a generalized convolution of the form (\ref{a9})
\cite{simu}.

In spite of the established formal correspondence, the interpretation and 
calculation of the components are entirely different. For instance, the
nuclear point-nucleon SF $f^{PN,A}$ in Eq. (\ref{a4b}) are calculated 
using characteristic nuclear tools and input, such as the single $N$ spectral 
function, $A$-particle density matrices of various orders, the effective 
$NN$ scattering amplitude, etc., whereas in Eq. (\ref{a8b}) those relate 
to the fsa $a(N$-Sp). Likewise, $F_2^N$ in Eq. (\ref{a1c}) is plainly taken 
from data, whereas in Eq. (\ref{a9}) it is the result of an elaborate 
pQCD calculation.  

We conclude this Section, emphasizing the different scales involved in 
the two factors of the integrand in Eq. (\ref{a9}), as has been illustrated
above on the example of quantum gases. In Eq. (\ref{a9}) by far the strongest
$Q^2$-dependence resides in $F_2^{\langle N \rangle}(x,Q^2)$, while 
the same in the nuclear component $f$ is soft. For $Q^2\gtrsim Q_0^2
\approx 3\,$ GeV$^2$ a parton description of the nucleon SF is largely 
sufficient, whereas the nuclear part including FSI, is most conveniently 
evaluated in a plain hadronic description. The above value $Q_0^2$ is 
approximately the one, above which Eq. (\ref{a1c}) has been found to hold.      

This concludes our generalization of the proofs of Jaffe and West on the
$'$factorization$'$ of nuclear pdf and SF. The next Section deals with
moments or Mellin transforms of nuclear SF in their obvious relation 
to $F_2^{\langle N \rangle}(x,Q^2)$.

\section{Moments of Nuclear Structure Functions.}
 
We recall the role played by moments  $M$ of $F_2^p$ for a $p$, for 
instance the Cornwall-Norton moments \cite{cn}
\begin{eqnarray}
M^p(n,Q^2)=(M_2^p(n,Q^2)=)\int_0^A dx x^{n-2}F_2^p(x,Q^2)
\label{a10}
\end{eqnarray}
For lowest twist (LO), non-singlets (NS) and large enough $Q^2$, asymptotic freedom 
of QCD predicts that moments of various rank raised to known powers are
linear in ln($Q^2)$. In terms of the strong coupling constant $\alpha_c$ 
\begin{mathletters}
\begin{eqnarray}
\label{a11}
\frac {M^p(n,Q^2)} {M^p(n,Q_0^2)} &\approx&
\bigg [\frac {{\alpha_c(Q_0^2)}} {\alpha_c(Q^2)}\bigg ]^{-d(n)}
\label{a11a}\\
\frac {{\alpha_c(Q_0^2)}} {\alpha_c(Q^2)}&\approx& 
1+\frac{\beta_0}{4\pi}{\alpha_c(Q_0^2)}
{\rm ln} \bigg (\frac {Q^2}{Q_0^2}\bigg ) 
+ {\cal O}\bigg ([\alpha_c(Q_0^2)]^2\bigg ),
\label{a11b}
\end{eqnarray}
\end{mathletters}
with $Q_0$ some scale and $\beta_0(N_f)=11-2N_f/3$ in terms of the number 
of flavors $N_f$. The exponents $d^{NS}(n,N_f)$ in Eq. (\ref{a11a}) are
expressed in terms of the NS anomalous dimension $\gamma_0^{NS}(n)$ 
\begin{mathletters}
\begin{eqnarray}
\label{a12}
d^{NS}(n,N_f)&=&\frac {\gamma_0^{NS}(n)}{2 \beta_0(N_f)}
\label{a12a}\\
\gamma_0^{NS}(n)&=&\frac {8}{3}\bigg [1-\frac {2}{n(n+1)}+4{\sum_ {2\le j\le n}}
\frac{1}{j} \bigg ]
\label{a12b}
\end{eqnarray}
\end{mathletters}
For conciseness we define
\begin{eqnarray}
{\cal S}^A&=& [M^A]^{-1/d(n)}
\nonumber\\
{\cal L}^A       &=&{\rm ln}({\cal S}^A)
\label{a98}
\end{eqnarray}
and find in view of  $|\frac {\beta_0}{4\pi}\alpha_c(Q_0^2)|\ll 1$
\begin{mathletters}
\begin{eqnarray}
\label{a13}
{\cal S}^p(n,Q^2) 
         &\approx & {\cal S}^p(n,Q_0^2)\bigg [1+\frac{\beta_0}{4\pi}
\alpha(Q_0^2) {\rm ln} \bigg (\frac {Q^2}{Q_0^2}\bigg )\bigg ]
\nonumber\\
         &=& c^p(n) {\rm ln}(Q^2) +b^p(n)\approx c^p(n) {\rm ln}(Q^2) +b^p
\label{a13a}\\
{\cal L}^p(n,Q^2) 
         &\approx& {\cal L}^p(n,Q_0^2)+\frac{\beta_0}{4\pi}
\alpha(Q_0^2) {\rm ln} \bigg ({Q^2/Q_0^2}\bigg )
\nonumber\\
         &\approx & \zeta^p(n) {\rm ln}(Q^2) +\eta^p(n)
\label{a13b}
\end{eqnarray}
\end{mathletters}
Slopes $c^p(n)$ for order $n$ and the common intercept $b^p$, are in principle 
determined by the scale or coupling constant, Eq. (\ref{a11b}). The 
predictions (\ref{a13a}), (\ref{a13b}) have decades ago been checked against 
available proton data \cite{penn}.  A recent JLab experiment, covering
$Q^2\lesssim  4.5 $GeV$^2$ and $x\lesssim x_M(Q^2)$ ($\approx$ 0.8 for
$Q^2=4.5\,$ GeV$^2$) allowed a more detailed analysis of the effects of
higher twist components in the moments $M^p(n,Q^2)$ \cite{osip}.

There has been hardly any interest in moments of nuclear SF \cite{rtval} 
for moderate  \cite{day} and large $Q^2$. In a straightforward way one 
can generalize the above for any target $A$, including for the averaged 
$N$, which requires in addition to $F_2^p$ knowledge of SF $F_2^n$, for which 
there is no direct experimental information. We refer to Ref. \cite{rt2} for 
the description of an indirect extraction of $F_2^n$ or $C(x,Q^2)=
F_2^n(x,Q^2)/F_2^p(x,Q^2)$ from inclusive scattering data on various 
targets. Once obtained,
\begin{eqnarray}
F^{\langle N \rangle}(x,Q^2)=\frac{Z+NC(x,Q^2)}{Z+N} F_2^p(x,Q^2)
\label{a14}
\end{eqnarray}
Since $F_2^n \ne F_2^p$, the parameter functions $c,b$ in (\ref{a13a}) 
for a neutron will differ from those for a proton and the same is the
case for the averaged $N$, or for any target $A$. In order to relate the 
latter two, one naturally exploits Eq. (\ref{a1c}) and its Mellin transform 
($m^{\langle N \rangle}\equiv 1$)
\begin{eqnarray}
M^A(n,Q^2)=m^A(n+1,Q^2) M^{\langle N \rangle}(n,Q^2)
\label{a15}
\end{eqnarray}
with
\begin{eqnarray}
M^A(n,Q^2)&=&\int_0^A dx x^{n-2}F_2^A(x,Q^2)
\nonumber\\
m^A(n,Q^2)&=&\int_0^A dx x^{n-2}f^{PN,A}(x,Q^2)
\nonumber\\   
\sigma^A(n,Q^2) &=&[m^A(n,Q^2)]^{-1/d(n)}
\label{a16}
\end{eqnarray}
A remark on $m^A$ is in order here. First, for $Q^2\gtrsim 20\,$GeV$^2$ 
one may neglect FSI parts in the calculated SF $f^{PN,A}$ from which $m^A$ 
is computed. Next, as moments of a peaked, normalized $f^{PN,A}$, 
$m^A(n=2,Q^2)$ has a minimum value 1, independent of $A$ and $Q^2$. For 
increasing $n$, $m^A(n)$ slowly increases, least for D, He and about to the 
same measure for all $A\gtrsim 12$. Those moments moreover carry the weak 
$Q^2$-dependence of $f$ \cite{rtv} and reach for $n$=7 the asymptotic 
limits $\approx 1.027$ for D and $\approx 1.082$ for medium and heavy A. 

For use below we also briefly discuss the behavior of $\sigma^A$, Eq.
(\ref{a16}). For $n$ between 2 and 7, the exponent $d(n,N_f=6)$ increases 
from 0.507 to 1.397 and causes $\sigma^A$ for D to barely decrease from 
1.000 to 0.997, and for $A\gtrsim 12$, from $1.000$ to $\approx 
0.931$. It suffices to illustrate (Fig. 4) the $n$-dependence of 
$m^A(n,Q^2=20\,$GeV$^2$) for D and Fe, representative for a target 
with $A\gtrsim 12$: The choice made for $Q^2$ is irrelevant, since the 
$Q^2$-dependence of $m^A$  is negligible for all practical purposes.

For target-independent anomalous dimensions, Eqs. (\ref{a14}), (\ref{a15}), 
(\ref{a16}) enable the generalizations of Eqs. (\ref{a13a}) and (\ref{a13b}) 
\begin{mathletters}
\begin{eqnarray}
\label{a17}
{\cal S}^{\langle N \rangle}(n,Q^2)
&\approx & c^{\langle N \rangle}(n) {\rm ln}(Q^2) +b^{\langle N \rangle}(n)
\label{a17a}\\
{\cal L}^{\langle N \rangle}(n,Q^2) &\approx& \zeta^{\langle N \rangle}(n)
{\rm ln}(Q^2) +\eta^{\langle N \rangle}(n)
\label{a17b}
\end{eqnarray}
\end{mathletters}
as well as
\begin{mathletters}
\begin{eqnarray}
\label{a18}
{\cal S}^A(n,Q^2)&\approx & c^A(n){\rm ln}(Q^2) + b^A(n)
\label{a18a}\\
{\cal L}^A(n,Q^2)&\approx & \zeta^A(n){\rm ln}(Q^2) + \eta^A(n)
\label{a18b}
\end{eqnarray}
\end{mathletters}
For given $Q^2$ we compared the separate expansions Eqs. (\ref{a18a}),
(\ref{a18b}) and found that the logarithm of the first is closely the  
second.

From the above one infers, that slopes and intercepts $c,b$ will differ 
for $p,n$ and thus for $\langle N \rangle$, while for general $A$ 
one checks from Eq. (\ref{a15}) the following approximation
\begin{mathletters}
\begin{eqnarray}
\label{a19}
c^{A}(2,Q^2)&\approx &\sigma^A(3,Q^2) c^{\langle N \rangle}(n)
\approx c^{\langle N \rangle}(n)
\nonumber\\
b^{A}(n,Q^2)&\approx&\sigma^A(n+1,Q^2) b^{\langle N \rangle}(n)
\approx b^{\langle N \rangle}(n)
\label{a19a}\\
\zeta^{A}(n)&\approx& \zeta^{\langle N \rangle}(n)
\nonumber\\
\eta^{A}(n,Q^2)&\approx&\eta^{\langle N \rangle}(n)+
   {\rm ln}\bigg (\sigma^A(n+1,Q^2)\bigg ) \approx \eta^{\langle N\rangle}(n)
\label{a19b}
\end{eqnarray}
\end{mathletters}
Medium changes are governed by $\sigma^A$, Eq. (\ref{a98}): target-to-target 
differences between slopes and intercepts for general targets and 
$\langle N\rangle$ never exceed a few $\%$ (see Fig. 4, the text after 
Eq. (\ref{a16}) and also point 5) below).
 
In what follows we distinguish between computed and experimental SF $F_2^A$ 
and their moments, as well as for ratios $\rho^A$, which derive from the 
Mellin transform (\ref{a15}) of the convolution (\ref{a1c}). For $M^{A,th}$ 
one has
\begin{mathletters}
\begin{eqnarray}
\label{a20}
\rho^{A,th}(n,Q^2)&\equiv& \frac {M^{A,th}(n,Q^2)}{m^A(n+1,Q^2)}= 
M^{\langle N \rangle}(n,Q^2) 
\label{a20a}\\
{[\rho^{A,th}(n,Q^2)]}^{-1/d(n)}&\approx& c^A(n,Q^2) {\rm ln}(Q^2)+b^A(n,Q^2) 
\nonumber\\
&\approx &c^{\langle N \rangle}(n,Q^2) {\rm ln}(Q^2)+
b^{{\langle N \rangle}}(n,Q^2),
\label{a20b}
\end{eqnarray}
\end{mathletters}
where we used Eq. (\ref{a15}). Clearly  $\rho^{A,th}(n,Q^2)=
M^{\langle N \rangle}(n,Q^2)$. For iso-singlet targets 
$M^{\langle N \rangle}(n,Q^2)$ does not depend on $A$, whereas for 
$I\ne 0$, there is a weak $A$-dependence due to the small neutron excess 
$\delta N/A$, Eq. (\ref{a1d}).

Using the measured $F^{A,dat}$, we consider the corresponding moments 
$M^{A,dat}$, Eq. (\ref{a16}), and the ratios $\rho^{A,dat}$, Eq. (\ref{a20}). 
In contrast to $\rho^{A,th}$, the ratios $\rho^{A,dat}$ do depend on 
$f^{PN,A}$.  A reliable computation of the latter and thus indirectly of
$m^A$, is presently only possible for $A \le 4$. 

Understanding the $n$-dependence of $M^A(n,Q^2)$ relies on the knowledge,
that all SF $F_2^A(x,Q^2)$ reach maxima for the smallest $x$, then 
decrease with increasing $x$ and become negligibly small beyond 
$x\approx 0.8$. The derived moments $M^A(n,Q^2)$ of lowest order thus 
critically depend on the values of $F_2^A(x,Q^2)$ for very small $x$ and 
on their accuracy. For growing $n,\,\, M^A(N)$ draws more and more 
from increasing $x$. Since for medium $x$, $F_2^A$ have fallen by at 
least an order of magnitude from their maxima, it becomes increasingly 
difficult to reliably compute $M^{A,dat}(n,Q^2)$  for large $n$. We now 
mention results for $N_f=6$.

1) $M^{A,th}(n,Q^2)$ is barely $A$-dependent, and for various $n$ slowly 
approaches its asymptotic $Q^2$ limit. In particular \cite{west,rt3}
$$M^{A,th}(n=2,Q^2\to \infty)\to \frac {5}{6}\frac{N_f}{(3N_f+16)}=0.1471$$

2) There is only meager experimental information available on $F_2^A$ for 
large $Q^2$. In spite of the fact, that second generation EMC ratios $\mu^A$
$=F_2^A/F_2^D$ have been measured for large $Q^2$, the individual $F_2^A$ 
are only rarely available. We know of CERN NA-4data on $F_2^A, A$=D, C 
\cite{dur,ben} and NA-2 for Fe \cite{dur,aub},
which are not dense and do not extend over the entire required critical 
$x$ range. To those we added a few data points from a JLab experiment 
\cite{arr3}, although the relevant $Q^2$ is low for a LO analysis. In
detail: The NA-4 $F_2^D$ data show substantial scatter \cite{dur}, which 
reflects 
in their moments and in ${\cal S}^D=[M^D]^{-1/d(n)}$. In spite of the 
above remarks, ${\cal S}^D$ for low $n$ accurately follows the theoretical
curves, but for increasing $n$, data overshoots predictions  up to 
$\approx 15 \%$ (Fig. 5). 

It is instructive to make a similar comparison for a large body of D 
data, which have been parametrized by Arneodo $et\,al.$ \cite{arn}. Very
good agreement now obtains for $n \le 4$. Discrepancies grow again 
with $n$, but are definitely smaller than for the above mentioned data
(Fig. 6). The cause is clearly  few percent differences between the two
data sets. The comparison also illustrates the effect of experimental scatter.

ii) The above data for $F_2^C$ lack values for small $x$ \cite{dur,ben} 
without which one cannot compute low-order moments. We therefore took 
recourse to a previously proposed method, which is based on the 
observation that $all$ $F_2^A(x\approx 0.18, Q^2)$ have a common value 
$\approx 0.30$, approximately independent of $A$ and $Q^2$ (see for instance 
Ref. \cite{rt2}). If nuclear SF are well-known for $x_m >x_0$, one may 
extrapolate $F^A$ down to $x_0$ (Fig. 7).  

iii) Before discussing Fe, we mention the result of a comparison of the
high-$Q^2$ data of Ref. \cite{dur,ben,aub} for $F_2^A(x,Q^2)$ 
a) $F_2^D \approx F_2^C$ 
b) For small $x$ both D and C are a few $\%$ lower than $F_2^{Fe}$, but  
for $x \gtrsim 0.18$ the situation appears reversed and $F_2^{D,C}$ are 
$\approx (15-20)\%$ larger than for Fe. No similar behavior has  been 
observed for lower $Q^2$. It is conceivable that the above Fe data \cite{dur} 
have a normalization error of the order of (15-20)$\%$ for $x \gtrsim 0.22$. 
In Fig. 8 we entered adjusted ${\cal S}^{Fe,dat}$. 

3) All ${\cal S}^{A,th}(n,Q^2)$ intersect around $Q^2 \approx (0.6-1.0)\,
$GeV$^2$, which is reflected in the approximate equality of all 
$b^A(n)$. The exception is $n=2$ for which ${\cal S}^A$ has a very 
small slope, which may reflect the sensitivity of $M^A(n=2,3)$ to the small 
$x$-behavior of the nuclear SF. The fact that there is little 
$A$-dependence seems to exclude screening effects in $F_2^A$ for 
$x\lesssim 0.15$ as a cause, but quarks emitted by  virtual bosons in 
the same small-$x$ range may contribute \cite{ls}.

4) Results for $\rho^{A,dat}$ and for $\rho^{A,\langle N \rangle;th}$
are assembled in Table I. There is overall agreement for D and C and 
a deficiency for the non-adjusted Fe data.

5) We tested, whether the expansions (\ref{a19a}) and 
(\ref{a19b}) for ${\cal S}^{A,th}$ and ${\cal L}^{A,th}$ and varying $n$, are 
approximately linear functions of ln($Q^2$) with only weakly $A$-dependent 
coefficients. Table II confirms the above for our targets. 

The  small, but marked influence of $m^A$ is manifest in a comparison
between${\cal S}^{\langle N \rangle}(n,Q^2)$ (for which 
$m^{\langle N \rangle} \equiv 1$) and ${\cal S}^A$. As Eq. (\ref{a19a}) 
predicts, intercepts $b^{\langle N \rangle}(n) \approx b^A(n)$ are quite 
similar, while for the slopes one has approximately 
$c^{\langle N \rangle}(n)\approx c^A(n=2)$. 

6) Finally we exploited the fact, that $\eta^A(n) 
\gg \zeta(n) {\rm ln}(Q^2)$ in Eq. (\ref{a18b}). Consequently
\begin{eqnarray}
\frac { {\cal L}^A(n,Q^2)} {{\cal L}^A(k,Q^2)} &\approx& 
\frac {\eta^A(n)}{\eta^A(k)}\bigg [1+ \bigg (\frac {\zeta^A(n)}{\eta^A(n)}
-\frac {\zeta^A(k)}{\eta^A(k)}\bigg ){\rm ln}(Q^2)\bigg ]
\nonumber\\
&\approx& \frac {\eta^A(n)}{\eta^A(k)}
\label{a21}
\end{eqnarray}
The bracketed form exceeds 1 by less than 10$\%$ and predicts only a weak 
$A$ and $Q^2$-dependence of the above ratios for pairs $n,k$, which gently 
grows with $n$-$k$. It thus suffices to illustrate the above for one species. 
We chose Fe and the pairs $n,k$=(4,2), (5,3), (6,4), (7,5), (7,3) (Fig. 9). 
The data are seen to follow the predictions (\ref{a21}) remarkably well, 
including the weak ln($Q^2$) dependence in (\ref{a21}). 

For a proton the linear dependence of ${\cal S}^p$ on ln($Q^2$) has been
regarded as support for asymptotic freedom. With quite similar results for 
nuclei, we do not tend to conclude the same for composite systems. It is 
more likely that the de facto separation of nuclear and nucleon components 
ascribes the above to the propagation of asymptotic freedom of (on-shell!) 
nucleons and allocates to the medium, controled  modifications of slopes 
and intercepts (see Ref. \cite{chen} for a differently argued separation). 

It is instructive to compare the above with an extention of the bag model 
of nucleons to nuclei with comparable average inter-nucleon spacings and  
sizes of bags, which may overlap and cause conceptual complications. No such 
problems occur in the above interpretation of the convolution (\ref{a1c}).  

Finally we remark that the above analysis is complicated by the presence of 
color singlet contributions, which are coupled to those for gluons. Only
for sufficiently high $n\gtrsim 4-5$ are those approximately decoupled
\cite{penn}, allowing an analysis of actual moments and not of the assumed 
non-singlets. This is also the reason, why we do not study medium
effects on slopes and intercepts in greater detail.

\section {Conclusion.}

The present note generalizes old work by Jaffe and West, who by means of a 
parton model and pQCD in the PWIA for large $Q^2$ proved that parton 
distribution functions and 
Structure Functions of composite targets and of nucleons are related 
by a generalized convolution. Their publications did not appear in 
the standard literature and have apparently been forgotten or disregarded. 
The present note is therefore in part an $amende\,honorable$ to their work.

We first reviewed facets of the conjectured convolution for $finite\,Q^2$, 
working in both a hadronic and an effective nuclear pdf representation. 
In those we did not aim to check, whether the $Q^2$ dependence is actually 
reproduced by, or in agreement with evolution from a scale $Q_0^2$. 

Next, we mentioned crucial points in the publications of Jaffe and West. 
Those are foremost the general relation between forward scattering 
amplitudes and parton distribution functions. Next we cited the
decomposition of fsa $a(\gamma$A) into the fsa $a(\gamma N$) and 
$a(N$,spectator-nucleus). Jaffe and West studied those first in the PWIA 
and in the Bjorken limit, leading to the scaling results. Those have 
subsequently been supplemented by contributions, due to gluon emission by 
quarks, etc. which, as regards photon-parton scattering, extend results  
beyond the above limit. We generalized the above and included classes of 
FSI between a nucleon, intermediately emitted by a target and the remaining 
spectator nucleus. The LO expressions, relating nuclear and nucleon pdf, 
and consequently the same for Structure Functions, continue to be of the 
convolution type. Moreover, those are identical to the same, previously 
conjectured ones in the above hadron and effective pdf representations. That 
correspondence is a formal one: the interpretation of the two results is 
entirely different.

The existence of an $'$ultimate$'$ description does not imply a preference 
over an $'$effective$'$ one under all circumstances. It is not only 
relatively easy to compute the  SF $f^{PN,A}$ from nuclear physics concepts 
than from pQCD, or to use data on nucleon SF, as opposed to a calculation of 
$F^N$: results from effective theories are frequently quite accurate.  

The above is not at all specific for descriptions of nuclear SF,
but holds for many $'$effective$'$ theories. A classical example 
is the inter-atomic interaction of the centers of the atoms in di-atomic 
molecules. The $'$true$'$ potential ought in principle to be derived from 
Schroedinger QM, which is extremely laborious, but in practice one uses
Lennard-Jones or Morse potentials. Those do contain the essentials of the 
physics, including a short-range repulsion, which mimics the effect of the 
Pauli principle for overlapping electron configurations. The spectroscopy of 
di-atomic molecules, and the physics of gases and liquids of 
di-atomic molecules is accurately accounted for by effective dynamics.

The last part of this note concerns moments of nuclear SF. The behavior 
of moments $M^p$ of the SF $F_2^p$, specifically the linear dependence of
${\cal S}^p$ on ln ($Q^2)$, has in the past been shown to be related to 
asymptotic freedom of QCD. Quite similar properties are shared by nuclear
moments. However, rather than concluding that inclusive scattering data on 
nuclei supports asymptotic freedom for composite systems, we prefer a
sober point of view. The formal factorization of $F_2^A$ (or the actual
one of moments $M^A(n,Q^2)$) separates nucleonic and nuclear dependencies, 
without changing the required separation of parts with hard and soft
$Q^2$-dependence as is the case of a proton. The observed ln$(Q^2)$ behavior 
simply reflects the propagation of asymptotic freedom of isolated nucleons, 
with characteristic medium modifications of nucleon parameters. 

\section {Acknowledgements.}
ASR thanks Pietro Faccioli and Marco Traini for assistance in initial
stages. Silvano Simula and Wally Melnitchouk provided expertise and critical
remarks.

\newpage

\begin{figure}[p]
\includegraphics[scale=1,angle=-90]{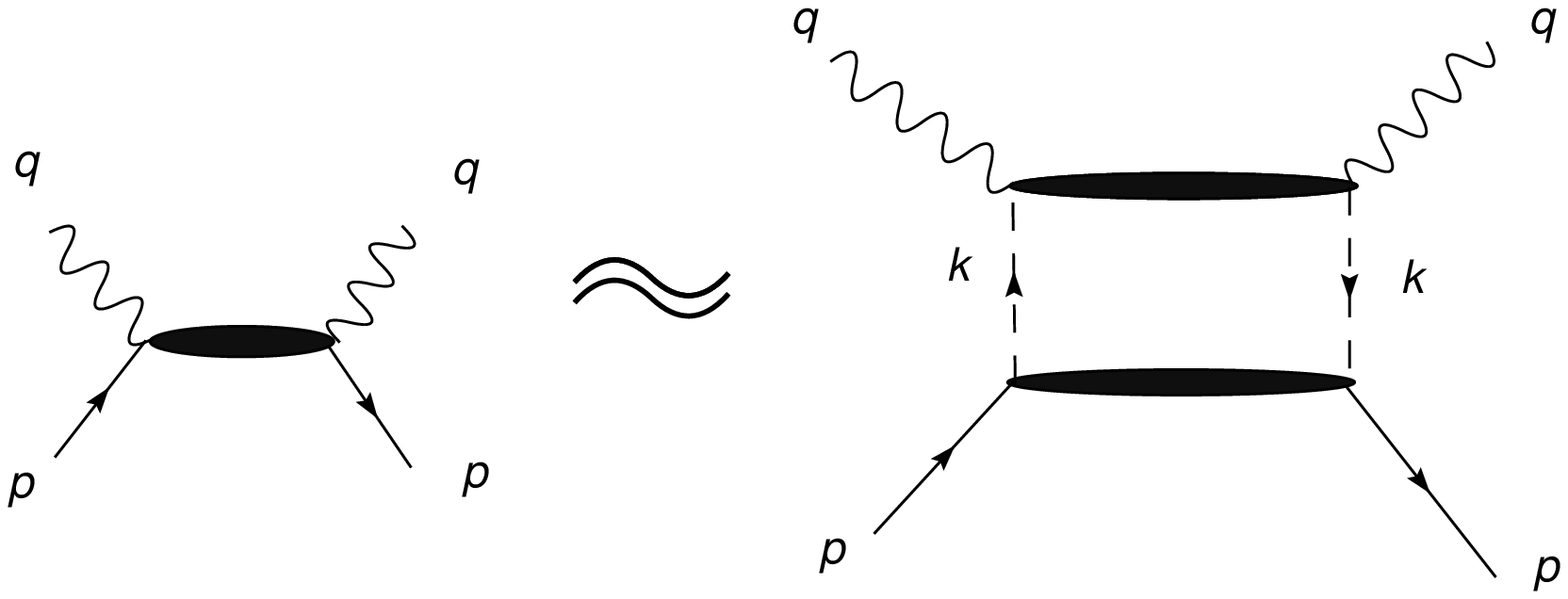}
\caption{ The decomposition of the forward $\gamma p$
amplitude in the PWIA and its link to the quark-$p$ scattering amplitude.}
\end{figure}

\begin{figure}[p]
\includegraphics[scale=.70,angle=-90]{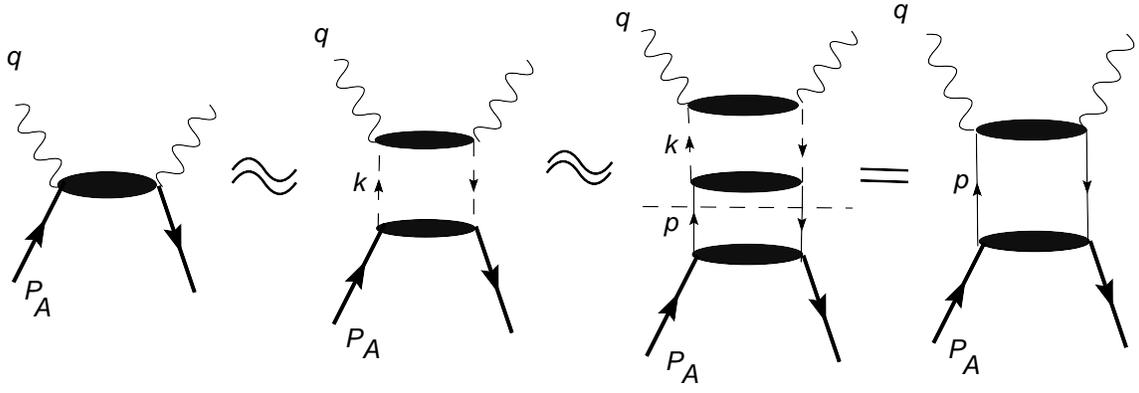}
\caption{ Same as Fig. 1 for a composite target. Inclusion of an 
intermediate set of free nucleon and spectator states, and recombination of
terms (marked by dashed horizontal), leads to a generalized convolution
of forward amplitudes $a(\gamma N)$ and $a(N$-Sp).}
\end{figure}

\begin{figure}[p]
\includegraphics[scale=0.70,angle=-90]{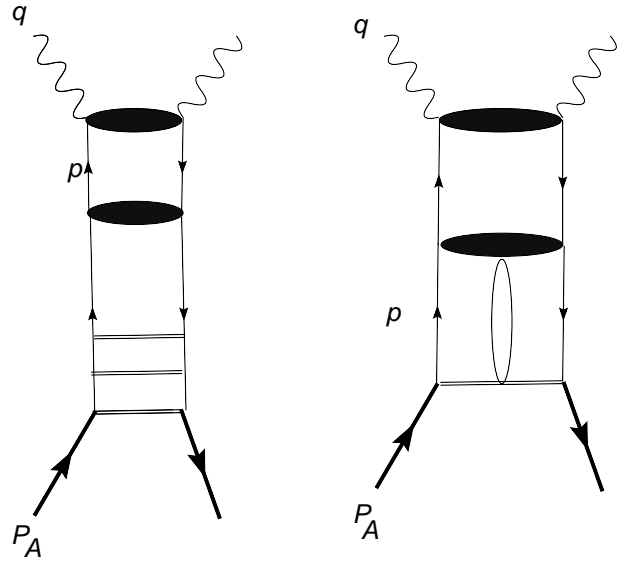}
\caption{ Ladder of $N$-Sp nucleus collisions, which are accommodated in
a convolution, and an example of nuclear FSI which cannot.}
\end{figure}
   
\begin{figure}[p]
\includegraphics[scale=1]{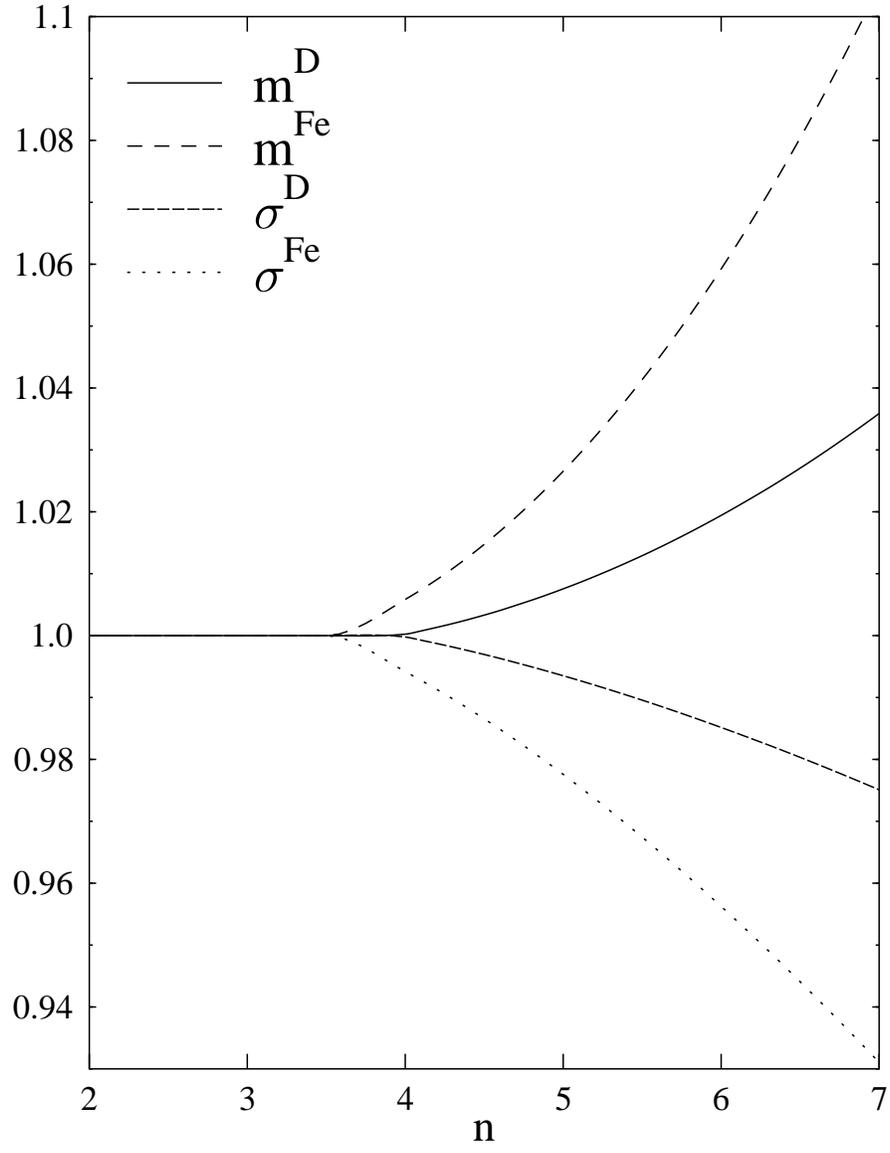}   
\caption{ Moments $m^A(n,Q^2)$ and its characteristic power $sigma^A(n,Q^2)$
Eq. (\ref{a13}), for $A$=D, Fe; $n=2-7$, $Q^2=20\,$GeV$^2$.}
\end{figure}

\begin{figure}[p]
\includegraphics[scale=1]{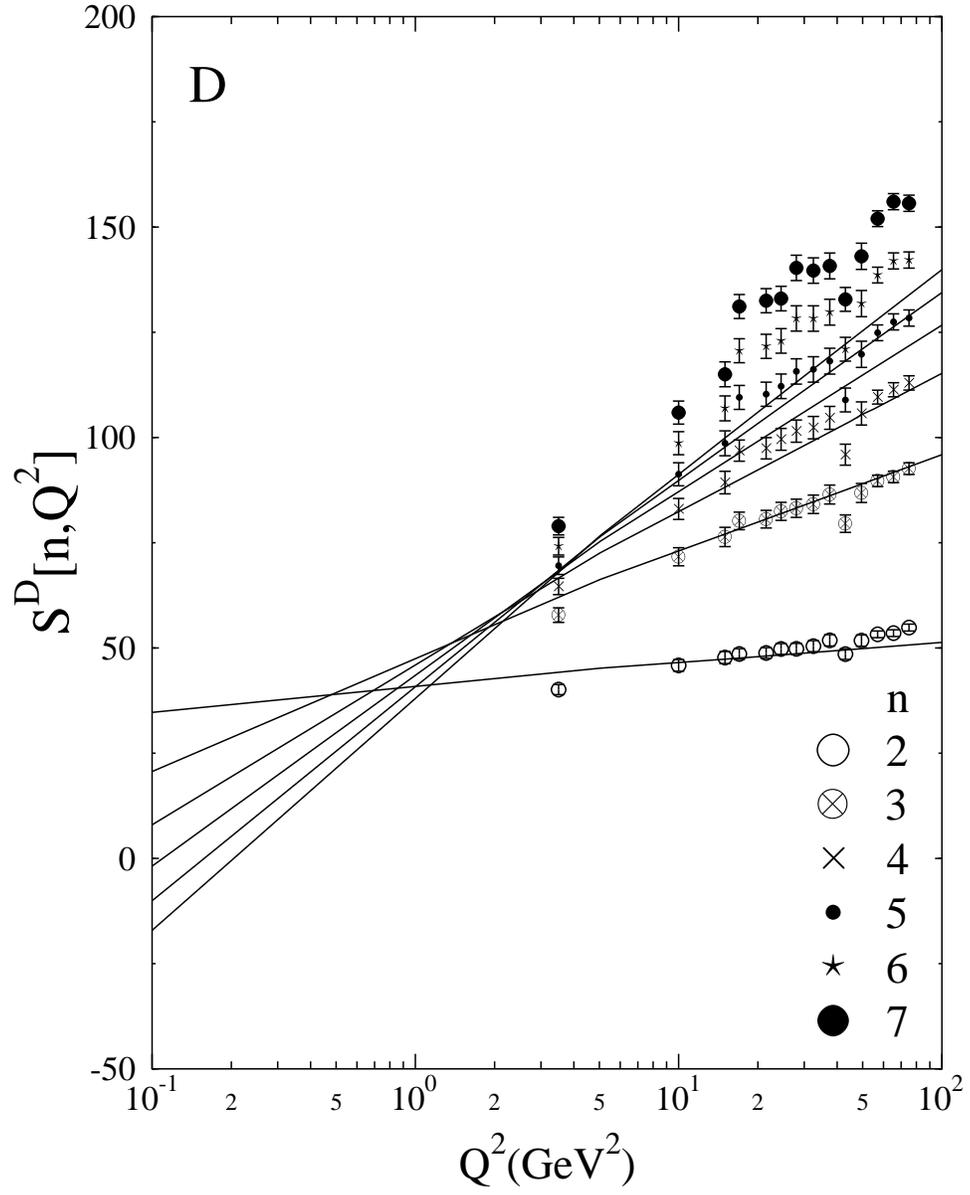}
\caption{
Characteristic powers of moments ${\cal S}^D(n,Q^2)$, Eq. (\ref{a98}),
for D as function of ln $Q^2$. Data points for underlying SF are
from Ref. \cite{dur} $n$ increases for lines with increasing slopes.}
\end{figure}                            

\begin{figure}[p]
\includegraphics[scale=1]{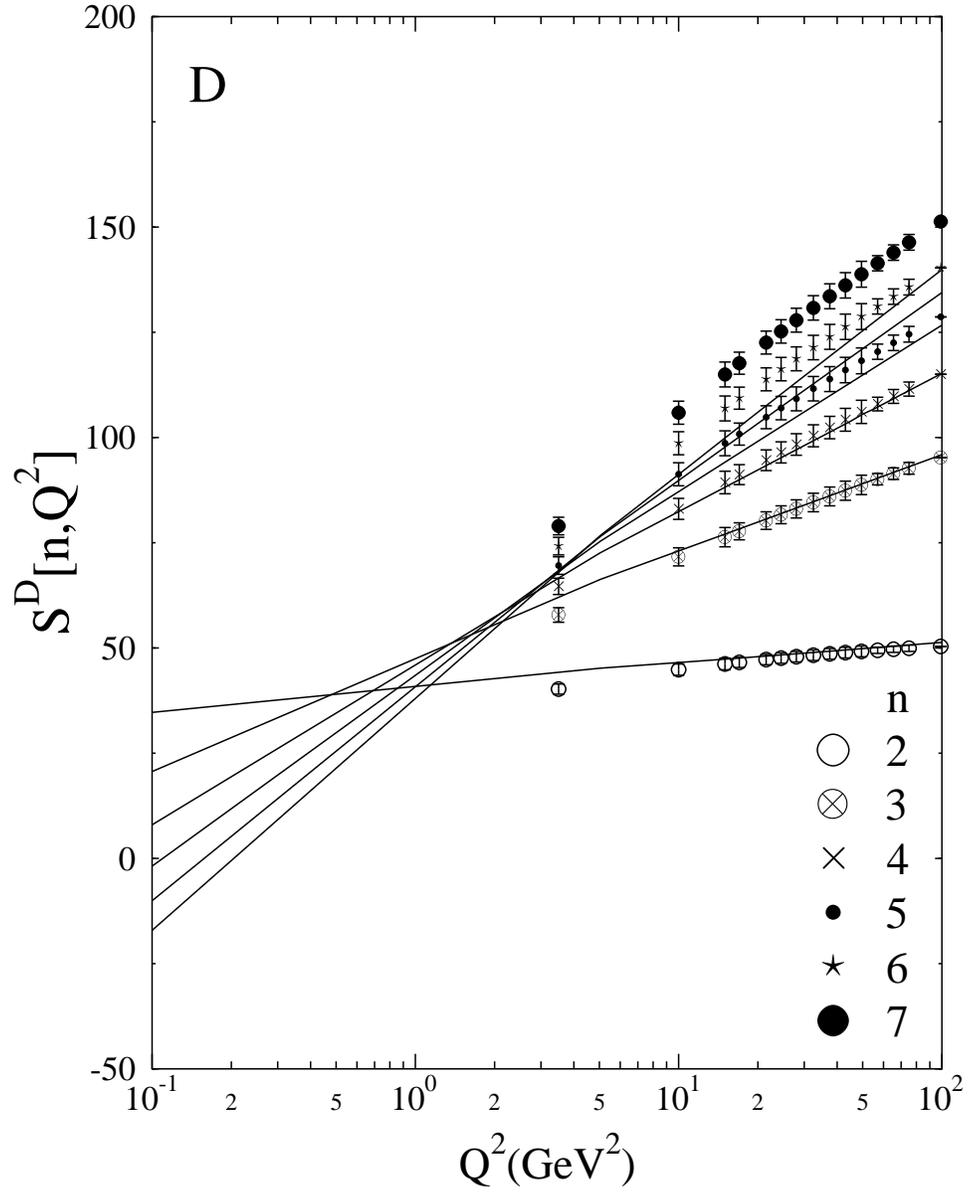}
\caption{                       
Same as Fig. 5 for parametrizations for the average of a vast body of
D data \cite{arn}.}         
\end{figure}

\begin{figure}[p]
\includegraphics[scale=1]{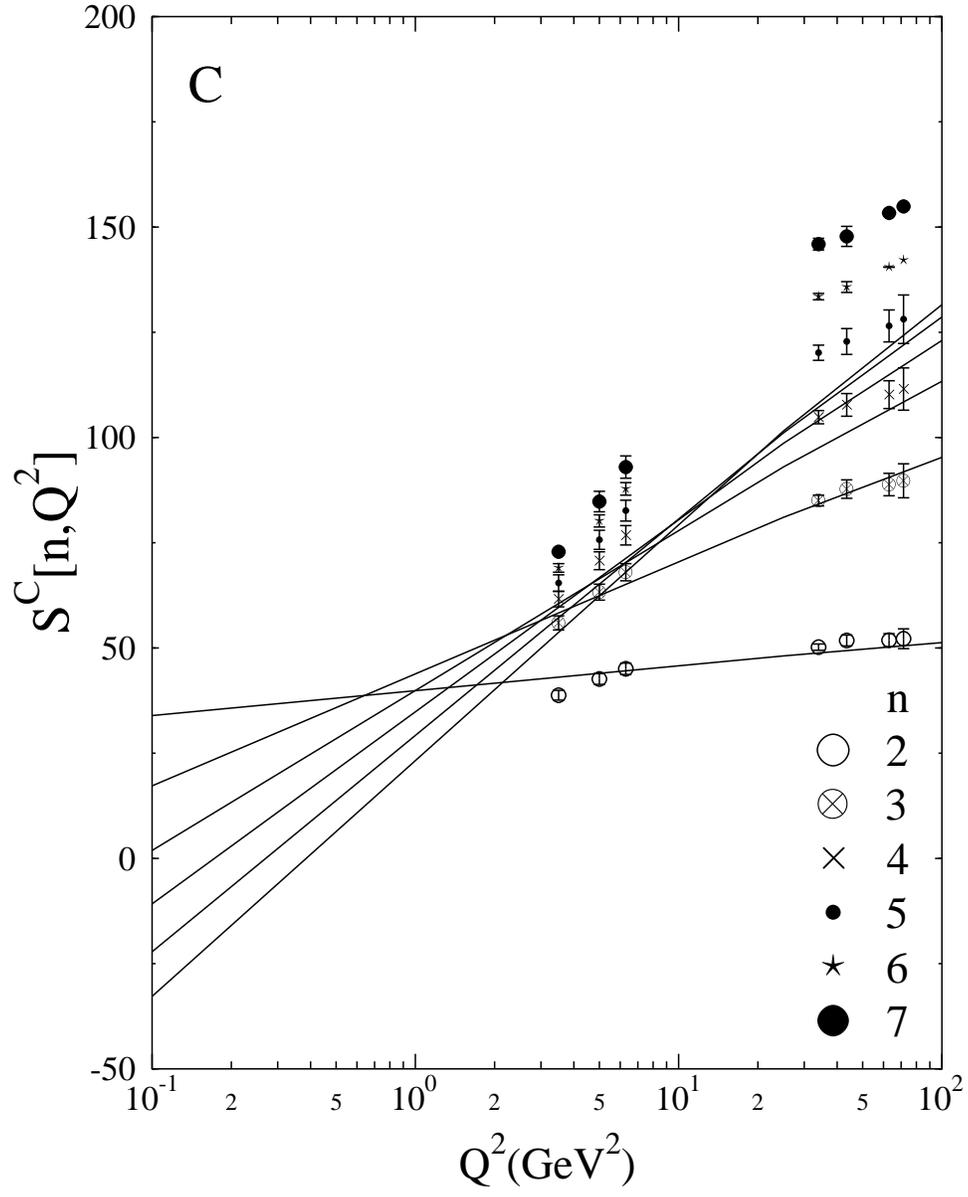}
\caption{ Same as Fig. 5 for C. Data are from Ref. \cite{dur,arr3}.}
\end{figure}

\begin{figure}[p]
\includegraphics[scale=1]{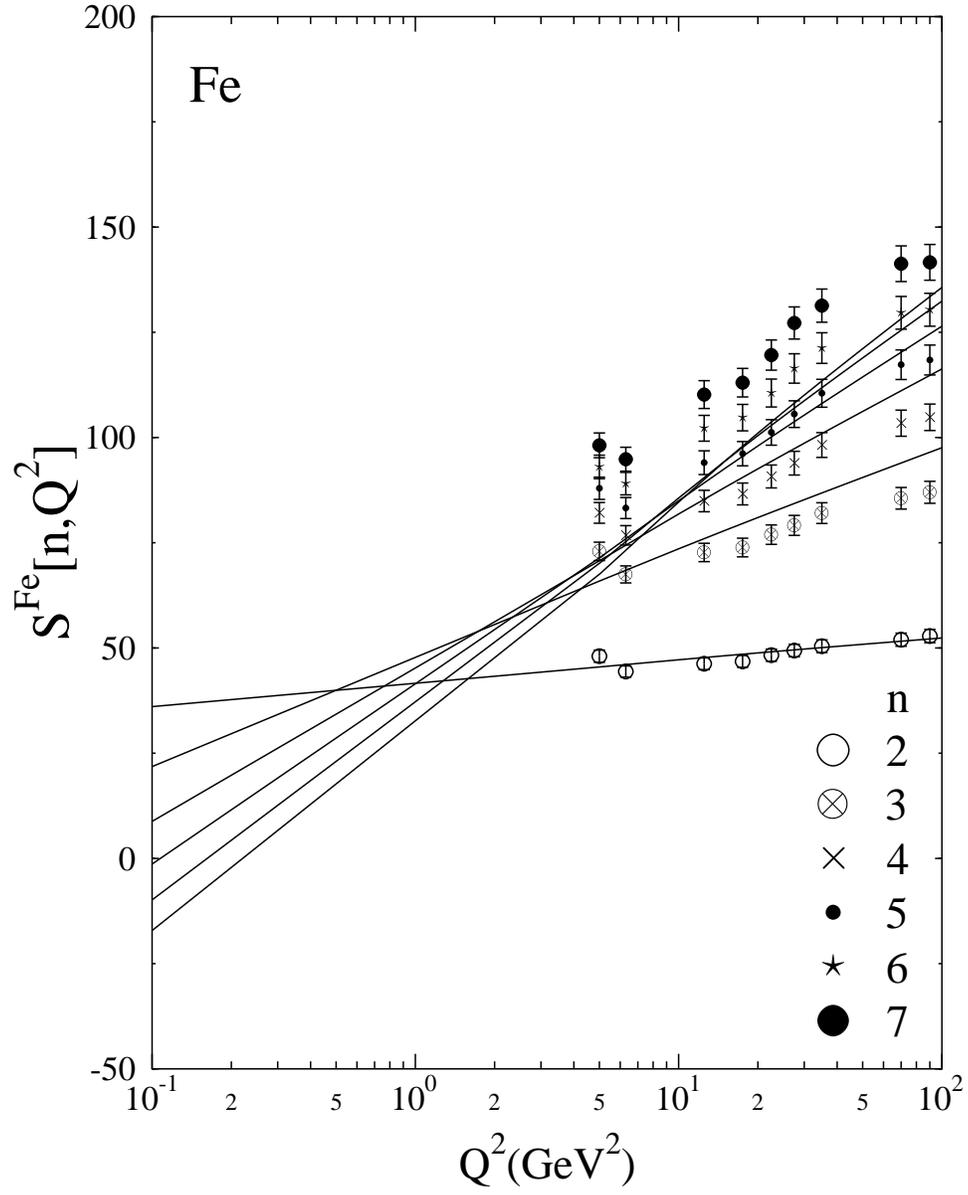}
\caption{ Same as Fig. 5 for Fe for partly renormalized $F_2^{Fe}$ data.}
\end{figure}

\begin{figure}[p]
\includegraphics[scale=1]{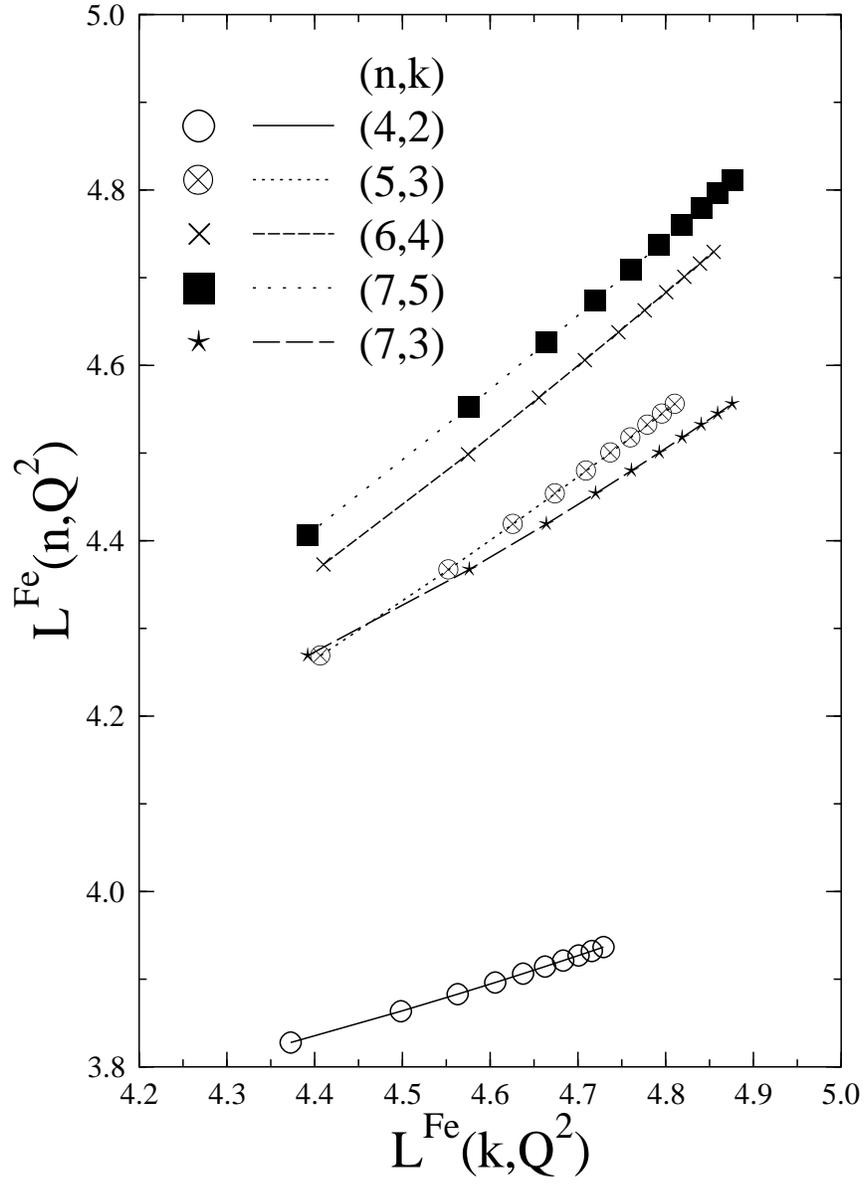}
\caption{ ${\cal L}^{Fe,th}(n,Q^2)$ versus ${\cal L}^{Fe,th}(k,Q^2)$, 
Eq. (\ref{a21}) for $(n,k)$= (4,2), (5,3), (6,4), (7,5), (7,3). Data points 
as in Fig. 7. }
\end{figure}

\newpage
\begin{table}
\caption {Ratios $\rho^{A,dat}(n,Q^2)$, Eq. (\ref{a20}), $A$= D, C, Fe,for
$n=2-7$ and
a number of roughly common $Q^2$ values.
Also shown are $\rho^{A,th}(n,Q^2)$ for the averaged $N$, pertinent to an
iso-scalar nucleus and for Fe.}

\begin{tabular}{|c|c|ccccc|}
\hline
target &~~ $n$~~ &\multicolumn{5}{c|}{$\rho^{A,dat}(n,Q^2)$} \\
\hline
$Q^2$ &    &   3.5&    17 &   35  &   50 &    72       \\
\hline
D~~~~~& 2  &~ 0.1548 ~&~ 0.1400 ~&~ 0.1372 ~&~ 0.1354 ~&~ 0.1315~
\vspace{-.5em} \\

~~~~~& 3  &~ 0.0401 ~&~ 0.0309 ~&~ 0.0298 ~&~ 0.0290 ~&~ 0.0276~ 
\vspace{-.5em} \\
~~~~~& 4  &~ 0.0156 ~&~ 0.0104 ~&~ 0.0099 ~&~ 0.0096 ~&~ 0.0090~ 
\vspace{-.5em} \\
~~~~~& 5  &~ 0.0073 ~&~ 0.0043 ~&~ 0.0040 ~&~ 0.0039 ~&~ 0.0036~ 
\vspace{-.5em} \\
~~~~~& 6  &~ 0.0038 ~&~ 0.0020 ~&~ 0.0019 ~&~ 0.0018 ~&~ 0.0017~ 
\vspace{-.5em} \\
~~~~~& 7  &~ 0.0021 ~&~ 0.0010 ~&~ 0.0010 ~&~ 0.0009 ~&~ 0.0008~ \\
\hline
C~~~~~& 2  & 0.1555 & 0.140  & 0.1373 & 0.1351 & 0.1347 \vspace{-.5em} \\
~~~~~& 3  & 0.0398 & 0.031  & 0.0291 & 0.0284 & 0.0281 \vspace{-.5em} \\
~~~~~& 4  & 0.0152 & 0.011  & 0.0094 & 0.0091 & 0.0089 \vspace{-.5em} \\
~~~~~& 5  & 0.0069 & 0.004  & 0.0037 & 0.0036 & 0.0035 \vspace{-.5em} \\
~~~~~& 6  & 0.0034 & 0.003  & 0.0017 & 0.0016 & 0.0016 \vspace{-.5em} \\
~~~~~& 7  & 0.0017 & 0.001, & 0.0008 & 0.0008 & 0.0008                \\
\hline
Fe~~~~& 2  & 0.1499 & 0.1325 & 0.1290 & 0.1270 & 0.1267  \vspace{-.5em} \\
~~~~~& 3  & 0.0354 & 0.0276 & 0.0260 & 0.0253 & 0.0251  \vspace{-.5em} \\
~~~~~& 4  & 0.0121 & 0.0087 & 0.0080 & 0.0077 & 0.0076  \vspace{-.5em} \\
~~~~~& 5  & 0.0049 & 0.0034 & 0.0030 & 0.0029 & 0.0029  \vspace{-.5em} \\
~~~~~& 6  & 0.0021 & 0.0014 & 0.0013 & 0.0012 & 0.0012  \vspace{-.5em} \\
~~~~~& 7  & 0.0009 & 0.0007 & 0.0006 & 0.0006 & 0.0006                \\
\hline
\hline
      &    &\multicolumn{5}{c|}{$\rho^{A,th}(n,Q^2)$}        \\
\hline
$\langle N\rangle_{I=0}$& 2  & 0.1469 &  0.1414 &  0.1393 &  0.1380 & 0.1369 ~
\vspace{-.5em}\\
~~~~~~~                & 3  & 0.0376 &  0.0315 &  0.0296 &  0.0285 &  0.0275 ~
\vspace{-.5em}\\
                    & 4  & 0.0149 &  0.0111 &  0.0100 &  0.0096 &  0.0092    ~
\vspace{-.5em}\\
                    & 5  & 0.0073 &  0.0050 &  0.0044 &  0.0041 &  0.0038    ~
\vspace{-.5em}\\
                    & 6  & 0.0041 &  0.0026 &  0.0022 &  0.0020 &  0.0019    ~
\vspace{-.5em}\\
                    & 7  & 0.0025 &  0.0015 &  0.0012 &  0.0011 &  0.0010    ~ \\

\hline
$\langle N\rangle_{Fe}$ & 2  & 0.1448 &  0.1396 &  0.1380 &  0.1374 & 0.1353 ~
\vspace{-.5em}\\                    
                    & 3  & 0.0368 &  0.0308 &  0.0295 &  0.0288 &  0.0272    ~
\vspace{-.5em}\\
                    & 4  & 0.0145 &  0.0109 &  0.0098 &  0.0094 &  0.0090    ~
\vspace{-.5em}\\
                    & 5  & 0.0071 &  0.0048 &  0.0044 &  0.0041 &  0.0037    ~
\vspace{-.5em}\\                    
                    & 6  & 0.0039 &  0.0025 &  0.0022 &  0.0020 &  0.0018    ~
\vspace{-.5em}\\                    
                    & 7  & 0.0024 &  0.0014 &  0.0012 &  0.0011 &  0.0010    ~\\
\hline
\end{tabular}

\end{table}

\begin{table}
\caption {Expansion coefficients of ${\cal S}^{\langle N\rangle,A;th}(n,Q^2)$,
Eqs. (\ref{a17a}), (\ref{a17b}), (\ref{a18a}, (\ref{a18b}) for $n=2-7$,
for $A$= D, C, Fe compared with the same for the average nucleon
$\langle N \rangle_{I=0}$.}

\begin{tabular}{|c|c|cc|cc|}
\hline
target
&~~$n$~~&~~$c^A(n)$~~&~~$b^A(n)$~~&~~$\zeta^A(n)$~~&~~$\eta^A(n)$        \\
\hline
D         ~& 2     &  2.085 & 46.503  &   0.0426 &   3.840 \vspace{-.5em} \\
         ~~& 3     &  9.937 & 73.035  &   0.1175 &   4.297 \vspace{-.5em} \\
         ~~& 4     & 14.273 & 82.415  &   0.1444 &   4.421 \vspace{-.5em} \\
         ~~& 5     & 17.717 & 87.177  &   0.1607 &   4.480 \vspace{-.5em} \\
         ~~& 6     & 19.379 & 89.868  &   0.1731 &   4.512 \vspace{-.5em} \\
         ~~& 7     & 21.111 & 91.332  &   0.1831 &   4.530                \\
\hline
C        ~~& 2     &  2.284 & 46.017  &   0.0469 &   3.830 \vspace{-.5em} \\ 
         ~~& 3     & 10.286 & 71.608  &   0.1232 &   4.278 \vspace{-.5em} \\ 
         ~~& 4     & 14.690 & 79.540  &   0.1526 &   4.386 \vspace{-.5em} \\ 
         ~~& 5     & 17.641 & 82.447  &   0.1723 &   4.425 \vspace{-.5em} \\ 
         ~~& 6     & 19.875 & 82.875  &   0.1891 &   4.433 \vspace{-.5em} \\ 
         ~~& 7     & 21.655 & 81.655  &   0.2049 &   4.420                \\
\hline
Fe       ~~& 2     &  2.241 & 47.245  &   0.0449 &   3.856 \vspace{-.5em} \\
         ~~& 3     & 10.404 & 73.705  &   0.1214 &   4.307 \vspace{-.5em} \\ 
         ~~& 4     & 14.917 & 82.088  &   0.1505 &   4.418 \vspace{-.5em} \\ 
         ~~& 5     & 17.937 & 85.366  &   0.1697 &   4.460 \vspace{-.5em} \\  
         ~~& 6     & 20.210 & 86.173  &   0.1856 &   4.471 \vspace{-.5em} \\
         ~~& 7     & 22.002 & 85.414  &   0.2002 &   4.465                \\
\hline
\hline
$\langle N\rangle_{I=0}$ & 2     &  2.137 &  40.989  &   0.0443 &   3.725 
\vspace{-.5em} \\
         ~~& 3     &  9.918 &  49.888  &   0.1182 &   4.020 \vspace{-.5em} \\
     ~~~~~~& 4     & 14.246 &  50.006  &   0.1443 &   4.093 \vspace{-.5em} \\
     ~~~~~~& 5     & 17.205 &  48.929  &   0.1598 &   4.126 \vspace{-.5em} \\
     ~~~~~~& 6     & 19.504 &  47.498  &   0.1711 &   4.144 \vspace{-.5em} \\
     ~~~~~~& 7     & 21.379 &  46.007  &   0.1800 &   4.154                \\
\hline
\end{tabular}

\end{table}

\end{document}